\documentclass[12pt]{iopart}
\usepackage{graphicx}
\usepackage{graphics}
\usepackage{floatflt}
\usepackage{float}
\usepackage{array}
\begin{document}
\title[Crystal nuclei and structural correlations in two-dimensional colloidal mixtures]
{Crystal nuclei and structural correlations in two-dimensional colloidal 
mixtures: experiment versus simulation}

\author{L Assoud\dag, F Ebert\ddag, P Keim\ddag, R Messina\dag, G Maret\ddag, H L\"owen\dag}

\address{%
\dag Institut f\"ur Theoretische Physik II: Weiche Materie,
Heinrich-Heine-Universit\"at
D\"usseldorf, Universit\"atsstra\ss e 1, D-40225 D\"usseldorf, Germany\\
}%
\address{%
\ddag Fachbereich f\"ur Physik, Universit\"at Konstanz, 
D-78457 Konstanz, Germany
}%
\date{\today}

\begin{abstract}
We examine binary mixtures of superparamagnetic
colloidal particles confined to a two-dimensional 
water-air interface both by real-space experiments and
Monte-Carlo computer simulations at  high coupling strength. 
In the simulations, the interaction is modelled 
as a pairwise dipole-dipole repulsion. While the
ratio of magnetic dipole moments is fixed, the interaction strength
governed by the external magnetic field and the relative composition
is varied. Excellent agreement between simulation and experiment is found for the partial
pair distribution functions including the fine structure of the 
neighbour shells at high coupling. Furthermore local
crystal nuclei in the melt are identified by bond-orientational order parameters
and their contribution to the pair structure is discussed.
\end{abstract}


\maketitle

\section{Introduction}

The mechanisms and principles of heterogeneous crystal nucleation
and the subsequent microstructure formation  are still
far from being explored and understood \cite{Auer1,Granasy1,Emmerich1}. 
Steering the nucleation behaviour has
important implications relevant for protein crystallization  \cite{Chayen} and
the formation of new glasses and metallic alloys \cite{Weinberg,Emmerich2}.
Colloidal suspensions have been exploited as model systems 
for crystal nucleation \cite{Palberg2,Kegel,Frenkel,Palberg1,Sear}. 
For instance, it is possible to watch crystal
nucleation in real space by using confocal microscopy \cite{Gasser}.
In particular, two-dimensional suspensions of superparamagnetic particles
confined to the air-water-interface of a pending droplet \cite{Keim07}
can easily be controlled by an external magnetic field. This allows
to tune the interparticle interactions and to quench the systems
quickly into a supercooled state \cite{Assoud_preprint}. The external field
induces magnetic dipole moments in the particles \cite{Zahn97}. If the field direction is normal
to the air-water interface, the resulting dipole moments are parallel 
and the interparticle pair potential is repulsive scaling with the inverse cube
of the particle separation \cite{Terada}. Binary mixtures
of these superparamagnetic colloids are ideal model systems to study 
crystal nucleation \cite{EPJE_Konstanz_2008}, crystallization \cite{Assoud_EPL}
and glass formation \cite{Konig} in real-space.

In this paper, we consider a binary mixture of superparamagnetic particles
both by real-space microscopy experiment and Monte-Carlo computer simulations of
a binary dipole-dipole interaction model \cite{Hoffmann,EPJE_Konstanz_2008}. 
In the simulations, the interaction is modelled 
as a pairwise dipole-dipole repulsion. While the
ratio of magnetic dipole moments fixed, the interaction strength
governed by the external magnetic field and the relative composition
is varied. 
We compare the pair correlation functions for
strong interactions (i.e. for large external magnetic fields) and find good agreement between
experiment and simulations. Moreover, we discuss the occurrence of peaks 
in the distance-resolved pair correlations in conjunction with local crystallites.
The latter are patches of triangular and square ordering which are crystalline
"seeds" in the amorphous fluid and building blocks from the globally
stable crystalline structure \cite{Assoud_EPL}. These crystallites actually could  act 
as nucleation centers for homogeneous nucleation if the system is quenched deeply
into the supercooled state \cite{Assoud_preprint}. 
Though heterogeneous nucleation is not tackled in the
present paper, an understanding of the local crystallites in the bulk is a first necessary step
in order to access possible pathways of inhomogeneous systems leading to
heterogeneous nucleation.

As regards to previous work, we here address strong couplings different from the weakly-coupled 
case where partial clustering of the small particles was found \cite{Hoffmann} and
equilibrated systems. In Ref. \cite{Assoud_preprint}, a quench was performed and the
pair correlation functions were found to be widely different from their
equilibration counterparts. For this genuinely non-equilibrium phenomenon, 
Brownian dynamics computer simulations were employed.

The paper is organized as follows: in section II briefly describe the experimental set-up
and the simulation method. In section III, results of experiment and computer simulations
are presented and discussed. Finally we conclude in section IV.

\section{ Methods}

\subsection{Experimental system and techniques}

The experimental system consists of a suspension of two kinds of spherical and
super-paramagnetic colloidal particles. The two species are called
 $A$ and $B$ respectively with $B$ referring to the smaller particles.
The hard-core diameters of the two species are $d_A=4.5\,\mu
m$, and  $d_B=2.8\,\mu m$ and their magnetic susceptibilities are
$\chi_A=6.2\cdot10^{-11}\:Am^2/T$ and 
$\chi_B=6.6\cdot10^{-12}\:Am^2/T$.  
The details of the experimental set-up are explained elsewhere
\cite{EPJE_Konstanz_2008,Keim07,Ebert_preprint_1}. Due to their high mass density, the particles are confined
by gravity to a flat water-air interface formed by a pending water drop.
The droplet is suspended by
surface tension in a top sealed cylindrical hole with a diameter of
$6\:mm$ and a depth of $1\:mm$ of a
glass plate. A coil produces a magnetic field $\textbf{H}$ perpendicular to the
water-air interface which induces a magnetic moment 
(i.e., ${\bf m}_i = \chi_i {\bf H}$ with $i=A,B$) 
in each particle. This leads to a repulsive dipole-dipole pair interaction \cite{Froltsov}.
By microscopy, trajectories of all particles in the field of view can be recorded
over several days providing sufficient phase space information. The
ensemble can be considered as ideally two-dimensional as the thermally
activated 'out of plane' motion of the particles is 
in the range of a few nanometer.

While temperature is fixed to room temperature the strength of the interparticle interactions
is tunable by the external magnetic field strength. 
A second parameter which is varied is the relative composition or the mixing ratio of the particles, 
%
\begin{equation}
X \equiv \frac{N_B}{N_A+N_B}
\label{eq:X}
\end{equation}
%

\subsection{Monte Carlo simulation technique}

In our Monte Carlo computer simulations,
we model the system in two spatial dimensions by a pairwise additive
potential 
%
\begin{equation} 
u_{ij}(r)=\frac{\mu_0}{4\pi} \frac{\chi_i \chi_j H^2}{r^3} 
\quad (i,j=A,B),  
\label{eq:u}
\end{equation}
%
where $r$ denotes the distance
between  two particles. For this inverse power potential, 
at fixed composition $X$, all static quantities
depend solely on a dimensionless interaction strength - or coupling constant -
%
\begin{equation} 
\Gamma=\frac{\mu_0}{4\pi} \frac{\chi_A^2H^2}{k_BTa^3}  
\label{eq:gamma}
\end{equation}
%
where $k_BT$ is the thermal energy
and  $a=1/\sqrt{\rho_A}$  the average interparticle separation between $A$-particles
\cite{Hansen_MacDonald_book}.
Hence effective temperature corresponds to the inverse of the coupling $\Gamma$, and the
system is completely characterized by three parameters: 
1) dipolar moment (or susceptibility) ratio 
%
\begin{equation} 
m \equiv \frac{m_B}{m_A} = \frac{\chi_B}{\chi_A},
\label{eq:m}
\end{equation}
%
2) the relative composition $X$, and 3) the interaction strength $\Gamma$.
While we fix the former to $m=0.1$,  the coupling $\Gamma$
and the relative composition $X$ are varied.
Standard Monte-Carlo simulations were performed with 
 $N_A=400$  $A$-particles and a corresponding number $N_B$ of $B$ 
particles determined by the prescribed relative composition $X$. The particles are
in a square box with periodic boundary conditions in both directions.
Typically $4\times10^6$ Monte Carlo steps per particle
are used for equilibration and statistics is gathered over additional $10^6$ Monte Carlo steps.

\section{Results}

We present our results for the two considered compositions $X=0.29$ and $X=0.44$.
Various coupling strengths $\Gamma$ are then investigated by (i) microstructural 
analysis and (ii) partial pair distribution functions.
In both cases, real-space experiments and Monte Carlo computer simulations 
have been performed.
As a reference, we have gathered on table \ref{table1} the different stable
crystalline structures at $m=0.1$ from the ground state ($T=0$) theoretical 
study  \cite{Assoud_EPL}.

%
\begin{table}[t]
\begin{center}
\caption{\label{table1}
      Theoretically predicted stable phases \cite{Assoud_EPL} for $m=0.1$ at $T=0$. 
      The same notation as in Ref. \cite{Assoud_EPL} is used here.
      The disks (open circles) correspond to $A$ ($B$) particles.}
{\setlength{\arrayrulewidth}{.3mm}
\begin{tabular}{|m{2.2cm}|m{1.8cm}|m{1.8cm}|m{1.8cm}|m{1.8cm}|m{1.8cm}|m{1.8cm}|}
\br
Phase& ${\bf T}(A)$ & ${\bf R}(A)A_3B$& ${\bf Re}(A)A_2B$ & ${\bf R}(A)AB$& ${\bf R}(A)A_2B_2$ & ${\bf S}(AB)$ \\
\hline
Composition ($X$)&0& 1/5       &  1/4       &    1/3   &   2/5      &  1/2   \\ 
\hline
&&&&&&\\
Crystalline structures &
   \begin{center}\includegraphics[width=1.6cm,angle=0]{fig_tab_a.eps} \end{center}&
   \begin{center}\includegraphics[width=1.6cm,angle=0]{fig_tab_b.eps} \end{center}&
   \begin{center}\includegraphics[width=1.6cm,angle=0]{fig_tab_c.eps} \end{center}&
   \begin{center}\includegraphics[width=1.6cm,angle=0]{fig_tab_d.eps} \end{center}& 
   \begin{center}\includegraphics[width=1.6cm,angle=0]{fig_tab_e.eps} \end{center}&
   \begin{center}\includegraphics[width=1.6cm,angle=0]{fig_tab_f.eps}   \end{center}\\
\br
\end{tabular}}
\end{center}
\end{table}
%

\subsection{Microstructural analysis}

\begin{figure}[t]
     \includegraphics[width=16.0cm,angle=0]{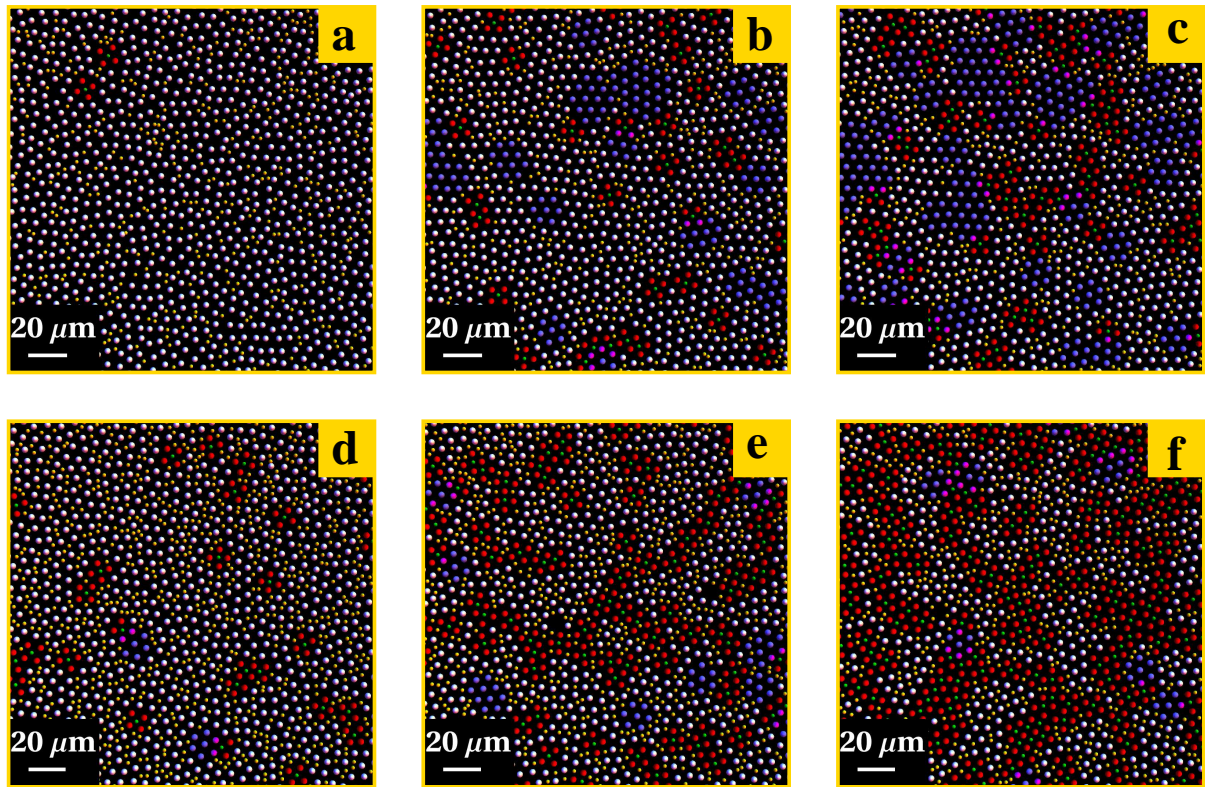}
     \caption{Experimental snapshots for the parameter combination 
       (a) $X=0.29$, $\Gamma=4.9$
       (b) $X=0.29$, $\Gamma=38.9$
       (c) $X=0.29$, $\Gamma=82.9$
       (d) $X=0.44$, $\Gamma=22.6$
       (e) $X=0.44$, $\Gamma=49.5$
       (f) $X=0.44$, $\Gamma=93.9$.
       Big particles are shown in blue if they belong 
       to a triangular surrounding and in red if the
       belong to a square surrounding. All other big 
       particles are shown in white color. 
       Few big particles belonging to both
       triangular and square surroundings are shown in pink color.
       The small particles are shown in green if they belong to 
       a square center of big particles,
       otherwise they appear in yellow.}
    \label{fig1}
 \end{figure}

A visual overview of typical microstructures from the experiments are provided in 
figure \ref{fig1}. 
From the theoretical study \cite{Assoud_EPL} (see also table \ref{table1}),
it is known that the relevant stable ground-state crystals consist of 
pure $A$-triangular [${\bf T}(A)$] structures ($X=0$) and intersecting squares of $A$ 
and $B$ particles [${\bf S}(AB)$ phase] at $X=0.5$.
Local crystallites in the fluid which possess this order are  detected by
coloring particles which have a pure triangular and square order.\footnote
 {  
 In detail, we have used criteria to define $A$ particles which have a pure triangular
 surrounding of other $A$ particles, 
 i.e.\ which are close to a cut-out of a pure triangular $A$ crystal, and,
 likewise, we have identified $A$ and $B$ particles which form locally an equimolar square lattice
 $S(AB)$. The corresponding two structure elements are shown in Table \ref{table1}. 
 In detail, we associate a triangular surrounding to an $A$ particle if the following two 
 criteria are fulfilled simultaneously
 \cite{EPJE_Konstanz_2008}: 
 1) The 6-fold bond order parameter $p_6 = \sqrt{\Psi_6^*\Psi_6}$
 (where $\Psi_6=\frac16 \sum_{NN}^6 \exp{{(i6\theta_{NN})}} $ with $\theta_{NN}$ denoting the
 angles  of the six nearest neighbour bonds relative to a fixed reference) is larger than $0.94$.
 2) The relative bond length deviation
 $b_6=\frac16 \sum_{NN}^6\frac{|l_{NN}-\bar{l}|}{\bar{l}}$
 where $\bar{l}$ is the average length of the six bond lengths $l_{NN}$ is smaller than $0.04$.
 This double condition selects local configurations close to those of a perfect
 triangular lattice where $p_6$ is unity and $b_6$ vanishes. Likewise we define a square 
 surrounding around a $B$ particle by the criteria:
 1) The 4-fold bond order parameter $p_4 = \sqrt{\Psi_4^*\Psi_4}$
 (where $\Psi_4=\frac14 \sum_{NN}^4 \exp{{(i4\theta_{NN})}} $ with $\theta_{NN}$ denoting the bond
 angles of the four nearest neighbour $AB$ bonds) is larger than $0.92$. 
 2) The corresponding relative $AB$ bond length deviation $b_4$ is smaller than $0.05$.
}

One can clearly see from figure \ref{fig1} that the triangular ${\bf T}(A)$
and square phases ${\bf S}(AB)$ are indeed predominant at strong enough couplings.
More precisely, at $X=0.29$, see figure \ref{fig1} (a-c),
there is a strong presence of triangular ${\bf T}(A)$-crystallites.   
This fraction of triangular crystallites is  growing with increasing $\Gamma$ 
see figure \ref{fig1} (a-c). 
%

At nearly equimolarity with $X=0.44$, see figure \ref{fig1} (d-f),
the situation differs qualitatively (compare with figure \ref{fig1} (a-c)), 
where we have now a strong fraction of squared ${\bf S}(AB)$-crystallites. 
This fraction is increasing with  growing $\Gamma$, 
which is consistent with the zero temperature limit
predicting the stability of the squared ${\bf S}(AB)$-lattice at $X=0.5$ 
(see table \ref{table1} and Ref. \cite{Assoud_EPL}).

\begin{figure}[t]
     \includegraphics[width=16.0cm,angle=0]{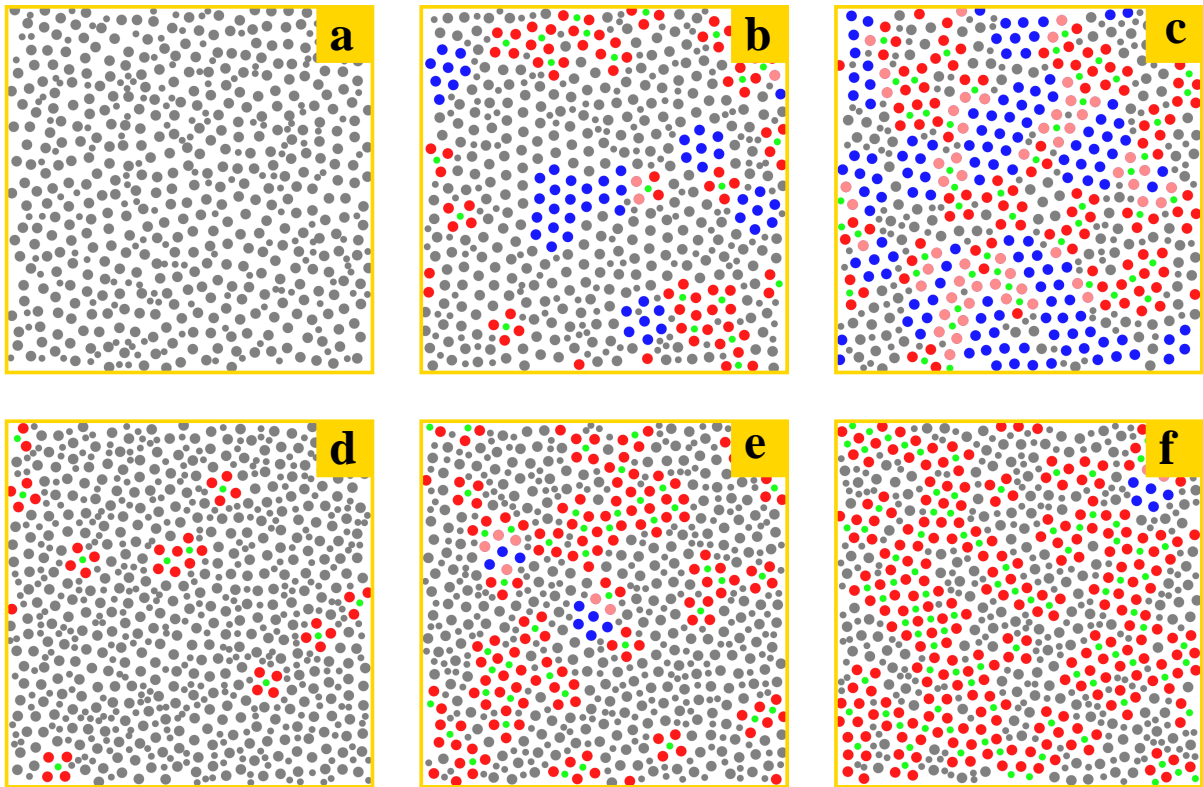}
     \caption{Simulation snapshots for the parameter combination 
       (a) $X=0.29$, $\Gamma=4.9$
       (b) $X=0.29$, $\Gamma=38.9$
       (c) $X=0.29$, $\Gamma=82.9$
       (d) $X=0.44$, $\Gamma=22.6$
       (e) $X=0.44$, $\Gamma=49.5$
       (f) $X=0.44$, $\Gamma=93.9$.
           Big particles are shown in blue if they belong 
           to a triangular surrounding and in red if the
           belong to a square surrounding. All other big 
           particles are shown in grey color. 
           Few big particles belonging to both
           triangular and square surroundings are shown in pink color.
           The small particles are shown in green if they belong to 
           a square center of big particles,
          otherwise they appear in grey.}
    \label{fig2}
 \end{figure}

The simulation snapshots are presented in figure \ref{fig2} for the same 
$(\Gamma,X)$ parameters as in figure \ref{fig1}.
In a general manner, there is an excellent qualitative agreement between the experimental and simulational
microstructures, compare figure \ref{fig1} with figure \ref{fig2}. 
Very interestingly, the theoretically predicted intermediate rectangular phase ${\bf Re}(A)A_2B$, 
see table \ref{table1} for $X=0.25$, is remarkably well present in the snapshot of 
figure \ref{fig2}(c). 
This feature was not detected in the experiments, compare with figure \ref{fig1}(c),
possibly due to a slightly imperfect equilibration thereby.

The emergence of crystalline clusters in the strongly interacting system gives some insight
into the nucleation behaviour. Presumably the system is in a stable crystalline phase at high $\Gamma$
but since it is undercooled it does not find the ultimate stable state \cite{Ebert_preprint_2}.
The intermittent crystal nucleation ``self poisons'' \cite{Schilling_Frenkel} further crystal growths
which may be similar to nucleation in liquid crystalline systems  \cite{Schilling_Frenkel}.
\subsection{Pair distribution functions}

\begin{figure}[t]
    \begin{center}
     \includegraphics[width=15.0cm,angle=0]{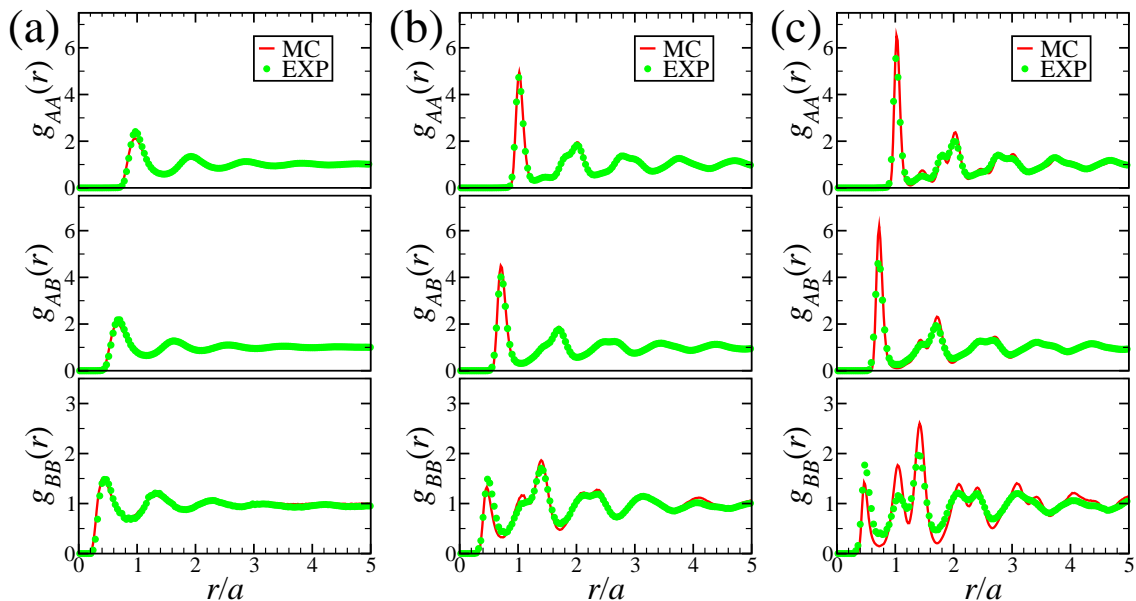}
     \caption{Partial radial pair-distribution functions 
       $g_{AA}(r)$,  $g_{BB}(r)$ and $g_{AB}(r)$. Experimental
       data (EXP) are compared to simulation results (MC) 
       for (a) $\Gamma=4.9$, (b) $\Gamma=38.9$ 
       and (c) $\Gamma=82.9$.
       The composition $X=0.29$ is fixed.}
    \label{fig3}
\end{center}  
 \end{figure}

We now discuss more quantitatively the structural aspects by inspecting
the radially averaged partial pair distribution functions, whose corresponding microstructures 
can be found in figure \ref{fig1} and figure \ref{fig2}
for the experimental and simulational data, respectively.

The case $X=0.29$ is reported on figure \ref{fig3}. 
In a general fashion, there is good quantitative agreement between 
experiment and simulation, see figure \ref{fig3}.
The only situation that slightly deviates from this quality of agreement,
concerns the partial pair distribution $g_{BB}(r)$ at $\Gamma=82.9$,
see figure \ref{fig3}(c). 
We explain this with a small drift of the colloidal system in the field of view. 
The whole systems contains more than 100000 particles and is susceptible to
perturbations. The drift induced some shear which may constrained here the ordering 
of the system at $\Gamma=82.9$ and $X=0.29$.
The primary peaks found at $r/a=1$ in $g_{AA}(r)$ and at $r/a= 1/\sqrt 2 \approx 0.71$ in $g_{AB}(r)$ 
is the signature of the squared ${\bf S}(AB)$-crystallites, see figure \ref{fig3}(b,c).

The disordered aspect of the material can be best identified by analyzing at  
the partial pair distribution $g_{BB}(r)$. 
Figure  \ref{fig3} (b,c) shows that the first peak in  $g_{BB}(r)$ is located
at $r/a \approx 0.45$, which is much smaller than the (square) unit lattice parameter, which is at $r/a=1$.
This is reminiscent to the similar trend for the small particles to form pairs and/or clusters 
at lower coupling \cite{Hoffmann}, 
as can also be easily observed on the microstructures of figures \ref{fig1} and \ref{fig2}.  

\begin{figure}[htb]
    \begin{center}
     \includegraphics[width=15.0cm,angle=0]{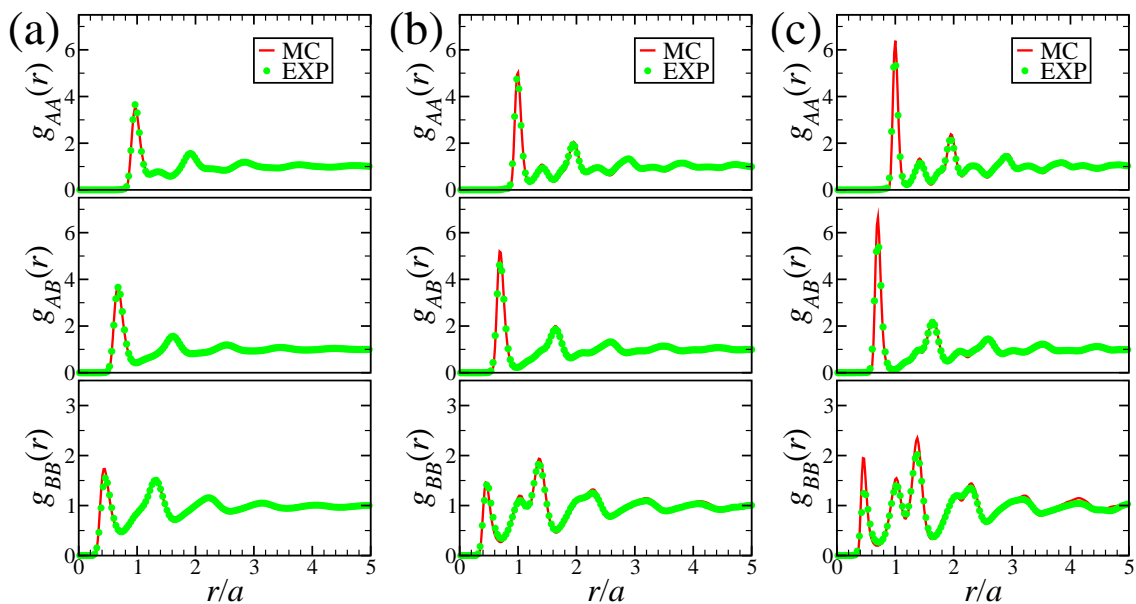}
      \caption{Partial radial pair-distribution functions 
           $g_{AA}(r)$,  $g_{BB}(r)$ and $g_{AB}(r)$. Experimental
           data (EXP) are compared to simulation results (MC) 
           for (a) $\Gamma=22.6$, (b) $\Gamma=49.5$ 
           and (c) $\Gamma=93.9$.
           The composition $X=0.44$ is fixed.
           }
     \label{fig4.eps}
   \end{center}
 \end{figure}

The case $X=0.44$ is reported on figure \ref{fig4.eps}. 
The agreement is now even better than at $X=0.29$,
becoming perfect for all the partial distribution functions
(including $g_{BB}(r)$).

\section{Conclusion}

In conclusion we have put forward the idea that two-dimensional binary
mixtures are excellent model systems for crystal nucleation as
they are easily supercooled by increasing an applied external field
and  crystallize into a variety of crystal structures. As revealed
by the excellent agreement in the pair correlation functions,
the system can be modelled by a simple dipole-dipole interaction 
potential \cite{Froltsov}.
The strongly interacting fluid bears some crystallites which were identified
and could act as possible nucleation centers both for homogeneous 
and heterogeneous nucleation.

Future investigations should consider the nucleation at fixed imposed nucleation
seeds which was studied theoretically for one-component two-dimensional systems
\cite{Sven}. Steering the nucleation and growth for binary systems is expected 
to be much richer since there are several competing crystalline structures.
Exploring more asymmetries in the magnetic moments is possible 
by exploiting nonlinear saturation effects in the magnetic
susceptibility at high external magnetic fields. This is another parameter which
is crucially determining the phase behaviour. Finally, it might be interesting 
to use binary charged suspensions confined between charged glass plates 
\cite{Murray_ARPC_1996,Lorenz_2008} as two-dimensional model system for crystal nucleation. The 
interactions are then well-approximated effective Yukawa potentials 
\cite{Yukawa_bulk,Yukawa_conf} where the screening length is steered by the salinity.
Again the ground state crystal structures  show a wealth of possible
crystals as recently revealed by lattice-sum calculations \cite{Assoud2}.
Therefore a rich scenario of crystal nucleation and growth phenomena 
are expected to occur here as well. Another realization of dipolar mixtures
in 2d are granular systems \cite{Hey_PRE_2003} which show interesting 2d ordering
effects \cite{Nelson_PMA_1982}.
Finally binary colloid mixtures
with added nonadsorbing polymers will result
in  effective attractions
and possibly liquid-gas phase separation \cite{Louis_JCP_2002,Rotenberg_MP_2004}.
The interplay of  vitrification or gelation
 and the fluid-fluid phase separation
 in two-dimensions should be an interesting topic for future research \cite{Zaccarelli_JPCM_2008}.

\ack
We thank D.R. Nelson, Y. Terada, and P. Dillmann for helpful discussions. 
This work was supported by the DFG (SPP 1296, SFB-513 project B6, and SFB-TR6 project C2).

\section*{References}


\end{document}